\documentclass{article}
\usepackage{calc}
\usepackage{graphicx}
\usepackage{amsmath}
\usepackage{amssymb}
\usepackage{relsize}
\usepackage{bm}

\usepackage[english]{babel}
\usepackage{multicol}

\usepackage{pgfbaselayers}
\pgfdeclarelayer{background}
\pgfdeclarelayer{foreground}
\pgfsetlayers{background,main,foreground}

\usepackage{times}
\usepackage{helvet}
\usepackage{bookman}
\usepackage{palatino}
\usepackage{float}
\usepackage{graphicx}

\selectcolormodel{cmyk}

\graphicspath{{images/}}

\newcommand*{\MAT}[1]  {\ensuremath{\mathbf{#1}}}
\newcommand*{\VEC}[1]  {\ensuremath{\bm{#1}}}

\date{December 10, 2013}
\begin{document}

\title{Contagion and Stability in Financial Networks} 

\author{
\vspace{1cm}\\
        Seyyed Mostafa Mousavi \\
              \small   Department of Statistics and Applied Probability, University of California Santa Barbara, CA \\\\
      Robert Mackay\\
       \small   Centre of Complexity Science, University of Warwick, Coventry, UK \\\\
      Alistair Tucker\\
       \small   Centre of Complexity Science, University of Warwick, Coventry, UK \\
       \vspace{1cm}\\
}

\maketitle

\begin{abstract}
This paper investigates two mechanisms of financial contagion that are, firstly, the correlated exposure of banks to the same source of risk, and secondly the direct exposure of banks in the interbank market. It will consider a random network of banks which are connected through the interbank market and will discuss the desirable level of banks exposure to the same sources of risk, that is investment in similar portfolios, for different levels of network connectivity when peering through the lens of the systemic cost incurred to the economy from the banks simultaneous failure. It demonstrates that for all levels of network connectivity, certain levels of diversifying individual banks diversifications are not optimum under any condition. So, given an acceptable level of systemic cost, the regulator could let banks decrease their capital buffers by moving away from the non-optimum area.
\end{abstract}
\section{Introduction}
The recent financial crisis has brought many scholars to the consensus that there exist some weak points in the current system and it is necessary to reform it to improve its stability (\cite{haldane2009rethinking}, \cite{french2010squam}, \cite{NYU}, \cite{IMFreport}). Safeguarding the stability of the whole financial system, or in other words, ensuring that the system works well even during bad economic periods in which some individual institutions may fail, is in charge of central banks. In fact, they are very concerned about the systemic risk, that is, the risk of a total system collapse like in 2007, after the failure of the Lehman Brothers (which was the counterpart of about \$5 trillion Credit Default Swap (CDS) contracts) there was a great fear that this failure might spread to the other institutions causing a cascade such that the whole economy could collapse. Also to avoid lending to distressed institutions, banks hoarded liquidity exacerbating the situation through financial contagion, that is when the solvency difficulties of some distressed institutions are spread like a disease to the other healthy ones. The culprit of financial contagion is the complex interconnections which exist between the financial institutions and generally, it crops up because of four mechanisms \cite{nier2007network}: 1) correlated exposure of banks to the same source of risk, 2) direct exposure of banks in the interbank market, 3) assets fire-sale of the bankrupted banks, 4) Informational contagion.\par

Correlated exposure of institutions to the same source of risk is when they invest in the same type of portfolio as the other institutions. According to the Modern Portfolio Theory, there is an optimum portfolio, which maximizes the expected return, for a given level of risk \cite{elton2009modern}. So, institutions, based on how risk-averse they are, would invest in their optimum portfolio and their portfolios would be very similar if they could tolerate almost the same level of risk. When individual institutions takes the long position on similar type of assets to maximize their individual return, the system become exposed to the risk that many banks fail simultaneously when those assets lose value for any reason. So, the regulator faces with a dilemma whether she has to let individual institutions maximize their return or she has to impose some regulations on the institutions to make them diversify their portfolio from that of each other \cite{BealePNAS}.\par 

At a moment in time, a bank may be in need of money for any reason, such as giving a loan to a customer or investing in an asset, while another one may have extra credit for which he does not have any investment plan, so a win-win decision could be the case when the second bank lends money to the first bank for a rate which is determined by the risk that the borrowing bank defaults on his loan. LIBOR, or London Interbank Offered Rate, is the interest rate at which banks can borrow funds from the other banks in London interbank market. When a bank fails, he might not be able to pay back his borrowings from the other banks and this would incur a loss to them and sometimes lead them towards bankruptcy. In bad economic states, banks become very fussy about lending in order not to incur any loss from the default of the borrowing bank and consequently, the interest rate soars or in other words, the credit dries up. For example, in the credit crunch of 2007, LIBOR surged from about 4\% to about 6\%. To name a few works investigating the interbank market as a source of contagion, it could be \cite{allen2000financial}, \cite{may2010systemic}, \cite{mistrulli2010assessing}, \cite{nier2007network}, \cite{gai2007contagion}, \cite{rochet1996interbank}, \cite{stiglitz2010risk} and \cite{wells2004financial}.\par

When a bank goes bankrupt, it has to follow the bankruptcy procedure and sell its assets to pay back his liabilities for as much as possible. Since the supply of those assets that the failed bank has in its balance sheet increases sharply, their prices tumble, so it is called the assets fire sale and it incurs great losses to those banks who have the same assets in their balance sheets. In fact, the bankruptcy of a bank can implicitly affect another bank which is not directly exposed to the failed bank at all (\cite{caballero2009fire}, \cite{boyson2010crises}, \cite{shleifer2010fire}, \cite{gai2007contagion} have discussed the effects of assets fire-sale).\par

Informational contagion happens when depositors or the other banks hoard liquidity even when the bank is working well and it causes a liability-side shock to the bank leading it to insolvency. In fact, it occurs because of the imperfect information about the bank assets and exposures. Historical evidence shows that there is a connection between dire liquidity problems and the economic state \cite{calomiris1991origins}. \cite{chen1999banking}, \cite{acharya2003information}, \cite{acharya2008information} and \cite{backus1999liquidity} are a few works in the literature in this regard.\par  

This paper is based on Beale et al paper \cite{BealePNAS}, which investigates the correlated exposure of banks to the same source of risk while quantifying the cost incurred to the economy from the banks simultaneous failure. In this paper, the direct exposure of banks in the interbank market, which could serve as a gate of contagion, has been also taken into account. This paper will consider a random network of banks which are connected through the interbank market and will discuss the optimum level of banks exposure to the same source of risk, that is investing in similar portfolios, for different levels of network connectivity when peering through the systemic cost lens. It demonstrates that for all levels of network connectivity, certain levels of regulators diversifying individual banks diversifications are not optimum under any condition. So, given an acceptable level of systemic cost, regulator could let banks decrease their capital buffers by moving away from the non-optimum area.\par

The rest of this paper is organized as follows. Section \ref{mathmodel} introduces the mathematical model. Section \ref{partition} discusses the partitioning of the assets loss space over possible configurations of banks state. Section \ref{simulationmodel} presents the simulation model. Section \ref{simresult} explains the simulations results and finally section \ref{conc} wraps up the conclusions.

\section{Mathematical Model}
\label{mathmodel}
\indent	In this model, a network of $N$ banks is considered. In this paper, bank is a general term and can refer to any type of financial intermediaries. Financial intermediaries can be classified as depository institutions (e.g. commercial banks), insurance and pension funds, finance companies, securities firms (e.g. investment banks and hedge funds) and mutual funds \cite{2}. Here, the same as some other works like \cite{gai2007contagion}, \cite{nier2007network} and \cite{haldane2011systemic}, each bank, as the node of the network, is schematically defined as follows:
\begin{figure}[H]
\begin{center}
   \includegraphics[width=0.4\textwidth,height=0.4\linewidth]{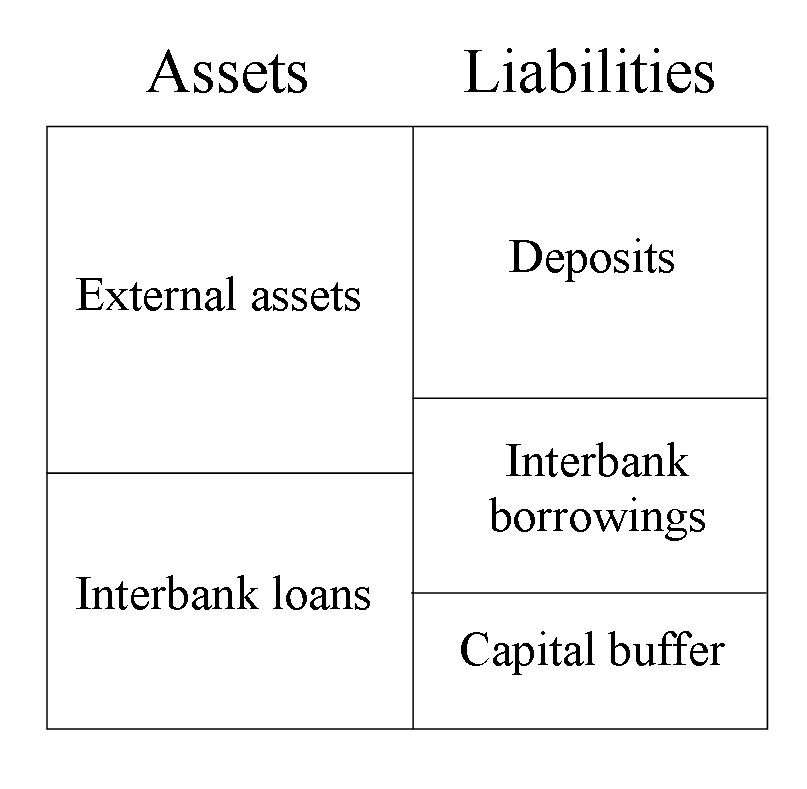}.
 \caption{Schematic view of a bank\label{fig:balancesheet}}
   \end{center}
   \end{figure}

\indent	For the "external asset" part of the asset side of the banks' balance sheet, each bank can invest in $M$ assets. The banks are connected to each other through interbank exposures, including interbank loans and interbank borrowings and they make a directional network. A bank would go bankrupt if it loses more than its net worth or its capital buffer for any reason, such as incurring loss from its investment portfolio, and then in this case it cannot return its liabilities to the other nodes of the system causing harm for them.\\
The mathematical model of the network is as follows:
   \begin{align}
      \VEC Y&=XV+LF        
    \end{align}
  $\MAT Y_i$: bank $i$'s asset side loss as a fraction of its capital buffer. \\
  \\
  $\MAT X_{ij}$: bank $i$'s investment in asset $j$ as a fraction of its capital buffer.\\
  \\
  $\MAT V_j$: simple loss of asset $j$ or how much asset $j$ has lost its value.\\
  \\
  $\MAT L_{ij}$: bank $i$'s loan to bank $j$ as a fraction of bank $i$'s capital buffer. Apparently, the elements on the diagonal of this matrix would be zero since a bank cannot lend money to itself.\\
  \\
  $\MAT F_i=I(Y_i\geq1)$: is 1 when bank $i$ loses more than its capital buffer or in other words is bankrupt. When the loss of bank $i$ is more than its capital buffer ($Y_i\geq1$), then the indicator function ($I$) would output the value of 1 as a result of bank $i$'s bankruptcy.\\\par
  Here the system is an excursive equation which should be initialized from a state, $F$ vector, which is usually the case when all banks are working well. Then, given the point of assets loss, $V$ vector, the final state of the system would be achieved from the excursion of the equation until the state does not change anymore with more excursion. From now on, we mean the final state with $F$ unless it is stated otherwise.\par
  It is worthwhile mentioning that there is no indication of time in this model. All banks' assets, including external assets ($X$ matrix) and interbank loans ($L$ matrix) are static not dynamic and the intuition is that this model is to understand how unstable different sorts of financial systems, $X$ and $L$ matrices, are from the regulator perspective and apparently this usually relates to the time of bad economic conditions when many assets lose value and cause loss to the holders and at that point, banks cannot evade the bad fortune by borrowing money from the other banks since the credit dries up and no other bank is keen to lend money to the distressed institutions for fear of increasing their possible loss as the case of those banks bankruptcy \cite{may2010systemic}.\par 
   As a corollary of the lack of time, an important tenet of the model is that it is contemplating the interbank loans only as a gate of contagion and is not taking into account its returns over time, so it is not the right means to analyze what level of interbank exposure is right for the system.\par
   
\section{Partition of assets loss space over the banks state}
\label{partition}
The $M$ dimensional space of assets loss vector, $V$, could be partitioned up to $2^N$ possible configurations of $F$, the binary vector of the banks state, and for the simple case of $M=N=2$, the partition diagram could be as follows:
\begin{figure}[H]
\begin{center}
   \includegraphics[width=1.0\textwidth,height=0.6\linewidth]{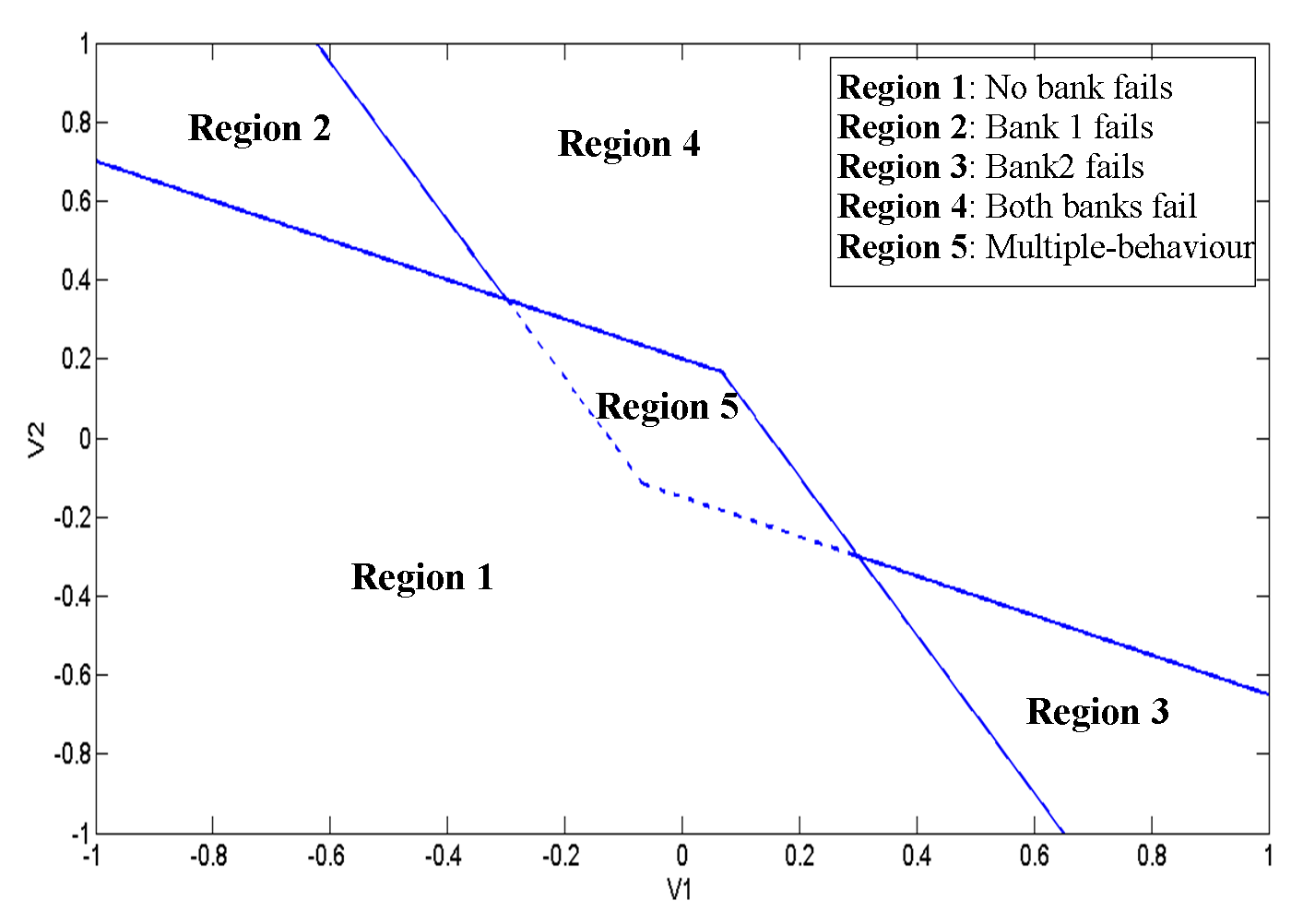}
  % \vspace{0.1cm}
   \caption{Partition of V-space over F\label{fig:partition}}
   \end{center}
\end{figure}
In the $V$-space, each bank, according to its investment portfolio in the assets, would make a hyperplane (a line in the 2-dimensional $V$-space) such that when the point of assets loss is higher than the hyperplane, the bank is bankrupt. Otherwise, it is working well. Each hyperplane would be displaced in the points where another bank over which the bank has claims has already defaulted because that bank cannot return its borrowings and consequently causes harm to the bank which has exposure to it. Consequently, it enlarges the regions on which the bank is liquidated to the extent of its exposure to the defaulted bank. In figure \ref{fig:partition}, the final bank lines which are from some point displaced according to the exposure of the bank to the other defaulted bank, are shown.\par

From the continuation of the shifted parts of each at least two hyperplanes, a multiple-behavior region would be made like region 5 in figure \ref{fig:partition}. The final state of this region is completely contingent on the state with which the system is initialized. For example, in region 5 of figure \ref{fig:partition}, if the system is initialized with the state in which both banks are working well, then in the final state, both banks would work well and if in the initial state, the banks are already liquidated, they would remain insolvent in the final state, so it has a multiple behavior tendency. In fact, this behavior could be characterized as the lack of confidence in the banks whose final state depends on the initial state. In other words, those banks are not in a very good state and are likely to go bankrupt soon. Apparently, any path in the $V$-space through which a few banks go bankrupt simultaneously would pass through a multiple-behavior region.\par

If the number of assets is greater than or equal to the number of banks, one could make an invertible coordinate change in the $V$-space such that all the hyperplanes become coordinate direction. In the new coordinate system, which is a linear combination of the old coordinate system, it is apparent that all $2^N$ possible configurations of $F$ would exist, so it can be said that if the number of assets is greater than or equal to the number of banks, then all $2^N$ possible regions would exist. Besides, there would be a multiple-behavior region between the hyperplanes of any at least two banks, which are mutually directly or indirectly (through another bank) exposed to each other, given a combination of the other banks state, so there would be up to $\sum_{i=2}^{N} {{N}\choose{i}}2^{(N-i)}$ multiple behavior regions. Interestingly, assuming that all banks are at least indirectly connected to each other, there would always exist a multiple-behavior region, one side of which all banks are well and on the other side, they are all insolvent. In other words, regardless of how banks invest in the assets and regardless of how they are all exposed to each other, some paths in the $V$-space are possible, through which all banks go bust simultaneously. \par
\section{Simulation Model}
\label{simulationmodel}
In order to do the simulation, a network of banks should be constructed and, a probability distribution is required over the $V$-space to numerically calculate the probability of the regions in the partition diagram of $V$-space over the configurations of $F$. Finally, a measure, which is called cost function here, should be defined so as to compare it for different parameters of the model. 

\subsection{Network Model}
In our network, the nodes represent the banks and node $A$ is connected to node $B$ if there is a link from node $A$ to node $B$ which shows how much bank $A$ has lent to bank $B$. Here the simplest type of random network, that is Erdos-Renyi model, is used. So, each node is connected to any other node with the fixed probability of $P$. \par
It is assumed that all banks are the same. In other words, their total assets are equal and they have the same capital buffer, $\eta$, that is represented as a percentage of the total assets. Consequently, they would diversify their external assets investment portfolio in exactly the same manner and the default of any bank incurs the same cost to the economy. \par   
In this model, how much each bank is willing to invest in interbank loans depends on the level of the connectivity of the network ($P$). When $P=0$, which means there is no link in the network, all banks invest only in external assets in the asset-side of their balance sheets. However, for the other values of $P$, all banks invest a portion of their total assets in interbank loans if it is possible (For any value of $P$ between 0 and 1, sometimes it is not possible for a particular bank to invest in interbank loans because it is not connected to any other bank and in this case it would invest only in external assets). The larger $P$ is, the more willing banks are to invest in interbank loans until the network is complete ($P=1$), and at that time all banks invest their highest percentage in interbank loans. The idea is that banks do not like to lend a large volume of money to an individual bank because of its high risk, but when the connectivity of the network increases, on average, a bank is connected with more banks, so, in fact, the bank has found the opportunity of lending to more banks and as a result, it would invest more in interbank loans. Besides, as far as it is assumed that all banks are the same, all banks divide their total interbank loans equally with the banks with whom they are connected and do not distinguish between them. \par
As \cite{BealePNAS} suggests, an important parameter regarding the investment portfolio of the banks in the external assets ($X$ matrix), is how much different banks diversify their portfolios from each other. This parameter, called $D$, could be calculated by the average distance between the asset allocations of each pair of banks as follows
\begin{align}
      \VEC D&=\frac{1}{2N(N-1)}\sum_{i}{\sum_{l\neq i}{\sum_{j}{|\acute{X_{ij}}-\acute{X_{lj}|}}}}       
    \end{align}
Here $\acute{X_{ij}}$ is $X_{ij}$, which is the investment of bank $i$ in asset $j$ as a fraction of its capital buffer, when it is normalized by $\sum_{j}{X_{ij}}$, which is the total investment of bank $i$ in the assets as a fraction of its capital buffer. When all banks invest in exactly the same portfolio, $D$ would equal to 0 and when they do not invest in any asset in which the others have already invested, $D$ would equal to 1.\par
For a given level of the connectivity of the network, it is desirable to generate random $X$ matrices with different values of $D$ and Appendix shows how to do that when the number of banks are equal to the number of assets. It is also of great importance to note that here it is assumed all assets are the same. In other words, banks are completely impartial to the assets and do not prefer any one to the others.\par 
\subsection{Probability distribution over $V$-space}
\label{sec:probability distribution}
For an asset, there are two types of return, including simple return and log return  and they are defined as follows:
\begin{align}
       \text{Simple return}&=\frac{P_f-P_i}{P_i} 			\\	
      \text {Log return}&=\ln\frac{P_f}{P_i}      
    \end{align}
Here $P_f$ is the final price of the asset and $P_i$ is the initial price of it. Log return could take any value while simple return has a bound that it cannot be less than -1 since asset price cannot be negative. When final price is close to initial price, these two returns would approximately be equal (\cite{bingham2004risk})\par

For simplicity, it is assumed that log return is distributed as $\frac{1}{\alpha}T$ when $T$ is a student $t$ distribution with 1.5 degrees of freedom and $\alpha$ is a critical value of $t$-distribution corresponding to the probability $q$ that a bank fails if it has invested only in that asset (the same probability distribution as \cite{bingham2004risk}). When the log return follows that distribution, the simple return would be distributed as $e^{\frac{1}{\alpha}T}-1$ so that its bound would not be violated. Here since the model deals with loss not return, it could be said that the loss is distributed as $1-e^{\frac{1}{\alpha}T}$. \par

In the model, what has been meant by assets is, in fact, asset classes, that is a group of assets whose returns/losses are correlated, but independent from that of the other asset classes. In other words, it is assumed that all asset returns/losses are independent of each other. Also because assets are assumed to be the same, it could be said that asset losses are independently and identically distributed as $1-e^{\frac{1}{\alpha}T}$.\par
\subsection{Systemic Cost}
There is a tension between the best investment portfolios for individual banks and the safest one for the system as a whole \cite{BealePNAS}. When banks individually try to diversify their portfolio to maximize their profit, they all invest in the same portfolio and as a result, they are all exposed to the same source of risk. This herding behavior is harmful for the stability of the system as a whole. From this aspect, since it is extremely difficult for the economy to absorb the costs incurred by the bankruptcy of several banks together, it is desirable that banks are never exposed to the same source of risk, so it would be less likely that some banks fail together. So, the regulator has  been faced with a dilemma whether to let the banks maximize their individual profits or to make some regulations for enhancing the stability of the system.\par
To explore this tension, the same notion of systemic cost as \cite{BealePNAS} is introduced. 
\begin{equation}
      \VEC Expected(C)=\sum_{i=0}^{2^N} P_iK_i^S
   \end{equation}
  $\MAT C$: systemic cost. \\
  \\
  $\MAT P_i$: Probability of region $i$ in the partition diagram of $V$-space over $F$.\\
  \\
  $\MAT K_i$: Number of failed banks in region $i$.\\
  \\
  $\MAT S$: Power which determines how nonlinear the cost is on the number of failed banks ($S \geq 1$).\par
  
\section{Simulation Results}
\label{simresult}
In order to develop the simulation, $P$ and $D$ both are discretized and for each pair of them, the average of the expected systemic cost over 1000 randomly generated $X$ and $L$ matrices is calculated. For calculating expected systemic cost, probability of each region in the partition diagram is required and that is done by using a 10000 $M$-elements $V$-tuples which are generated from the probability distribution introduced in section \ref{sec:probability distribution}. \par
The values of the model parameters which are used in these simulations are summarized in table \ref{Parameters} although the results are still valid for the other choice of parameters.\par
\begin{figure}[H]
\begin{center}
   \includegraphics[width=1.0\textwidth,height=0.85\linewidth]{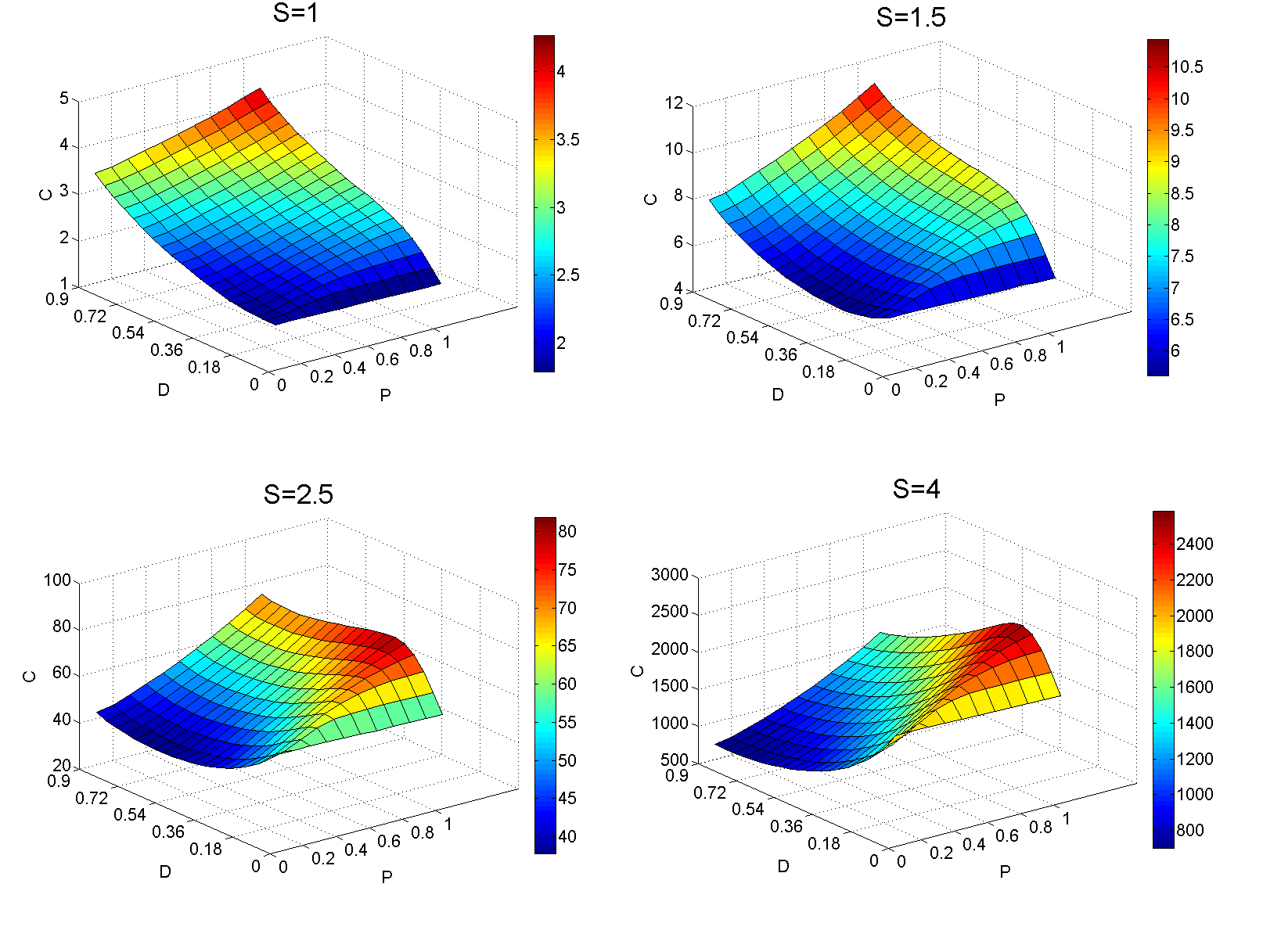}
  % \vspace{0.1cm}
   \caption{Average of the expected systemic cost versus network connectivity level and the diverse diversification parameter for four degrees of nonlinearity of the cost function\label{fig:mixeds}}
   \end{center}
\end{figure}

\begin{table}[h]
\begin{center}
    \begin{tabular}{ | p{9cm} |  p{1cm}| }
    \hline
    Parameter & Value \\ \hline
    Number of banks & 10 \\ \hline
    Number of assets & 10 \\ \hline
    Maximum Percentage of interbank loans to total assets (when P=1) & 20\% \\ \hline
    Type of the function of percentage of interbank loans to total assets  versus connectivity  & Linear \\  \hline
    \end{tabular}
\end{center}
\caption{Summary of the parameters of the model}
\label{Parameters}
\end{table}
The average of the expected systemic cost in figure \ref{fig:mixeds} is plotted against network connectivity, $P$, and diverse diversification parameter, $D$, for different nonlinearity levels of the cost on the number of failed banks, $S$. \par
 
%\begin{figure}
 % \centering
 % \subfloat[A gull]{\label{fig:gull}\includegraphics[width=0.3\textwidth]{s1}                
 % \subfloat[A tiger]{\label{fig:tiger}\includegraphics[width=0.3\textwidth]{s25}}
 % \subfloat[A mouse]{\label{fig:mouse}\includegraphics[width=0.3\textwidth]{s4}}
  %\caption{Pictures of animals}
  %\label{fig:animals}
%\end{figure}
%\label{myfigure}
%\caption{Global figure caption}
%\end{figure}

For different levels of network connectivity, it is interesting how $D_{opt}$, $D$ in which the cost is minimum, changes with the increase of $S$. When $P=0$, there is no interbank linkage, and also the cost function is linear ($S=1$), $D_{opt}$ equals to 0 since there is no premium cost for the case that some banks fail together and consequently, the systemic cost is minimum when all banks invest in their same optimum portfolio. However, when $S$ increases, $D_{opt}$ shifts away gently from 0 towards 1 depending on how nonlinear the cost function is, and this means banks would diversify their diversification to minimize the extra cost associated with the simultaneous failure of the banks. This is in harmony with the results in \cite{BealePNAS}\par
When $P>0$, that is when banks to some level are connected to each other, the pattern of change in $D_{opt}$ by the increase of $S$ is different from that of the case when $P=0$ since the interbank linkages can be served as a gate of contagion. Here $D_{opt}$, at first, remains constant at 0 when $S$ increases, but there exists a brink point in $S$ at which $D_{opt}$ swiftly changes from 0 to some value between 0 and 1. The higher the connectivity of the network is, the bigger this jump is and the later the brink point would happen. After this sudden change in $D_{opt}$, it would again drift gently towards 1 as $S$ increases.\par
\begin{figure}[H]
\begin{center}
   \includegraphics[width=0.5\textwidth,height=0.5\linewidth]{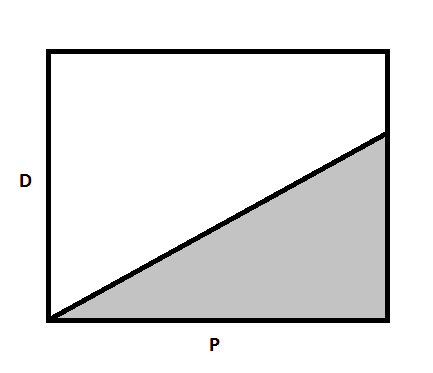}
  % \vspace{0.1cm}
   \caption{Non-optimum area in $P-D$ space. The gray area could not be optimum under any form of cost function\label{fig:area}}
   \end{center}
\end{figure}

In fact, there is an area in $P-D$ space in which the systemic cost would never be minimum and it is the area over which $D_{opt}$ jumps and is schematically shown in gray colour in figure \ref{fig:area}. From system stability point of view, the system should not be in that area under any form of the cost function since given any level of network connectivity, there always exists another $D$ out of that area in which the systemic cost is lower. In fact, the regulator could let banks decrease their capital buffers when the system is moving away from the non-optimum area while the systemic cost still remains at its acceptable level.\par  

The intuition behind the existence of such an area in $P-D$ space is that when banks are connected to each other, interbank loans would work as a gate of contagion, but when $D=0$, or in other words all banks have invested in the same portfolio, interbank loans could not spread the difficulty of distressed institutions to the other banks since when one bank has failed, it means that all the other banks are already bankrupt (because they have invested in exactly the same portfolio). So, even for very nonlinear cost functions, as $D$ drifts from 0, the systemic cost would increase at first because this diverse diversification would open the gate for interbank linkages to spread financial difficulty. However, after some point, for nonlinear enough cost functions, the systemic cost would decrease because of having less of the cost associated with the simultaneous failure of banks. In fact, this phase transition in $D_{opt}$ occurs because of this initial increase in the systemic cost as $D$ moves from 0 to 1 for any type of cost function.\par 

It is important to mention that the results do not depend on these particular parameters and by changing the parameters, including number of assets, number of banks, banks' capital buffer, maximum percentage of interbank loans to total assets, and type of the function of percentage of interbank loans to total assets versus connectivity, the pattern of the results remains intact. Besides, the results are not because of this particular assets loss probability distribution. If the t-distribution be swapped with another distribution like the normal distribution, the same type of results would be still achieved.\par

\section{Conclusion}
\label{conc}
This paper extends Beale et al model in \cite{BealePNAS} to account not only for the correlated exposure to the same source of risk, but also for the direct exposure of banks in the interbank market. We found that the existence of interbank loans which could serve as a means of contagion makes certain levels of the regulator diversifying the banks individual diversification non optimum given a level of network connectivity for any type of systemic cost function. Also, this moral for the banking system does not depend on the input parameters, including number of assets, number of banks, banks' capital buffer, maximum percentage of interbank loans to total assets, and the type of the function of percentage of interbank loans to total assets versus connectivity. So, in fact, the regulator could let banks decrease their capital buffers when the system is moving away from the non-optimum area while the systemic cost still remains at its acceptable level.\par

\bibliographystyle{abbrv}
\bibliography{project}
\renewcommand{\theequation}{A-\arabic{equation}}
  % redefine the command that creates the equation no.
  \setcounter{equation}{0}  % reset counter 
  \section*{Appendix: \\
  Randomly generating X matrices with a specific diverse diversification parameter}  % use *-form to suppress numbering
  Here is the algorithm that we have devised to randomly generate $X$ matrices with a specific diverse diversification parameter, $D$, given that the number of assets equals to the number of banks ($M=N=n$) or in other words, the $X$ is a square matrix.\par
  At first, a seed point, consisted of $n$ numbers, is randomly generated such that the sum of the $n$ numbers equals to 1. When $D=0$, all banks invest in the same type of portfolio as the seed (in fact, they multiply the seed according to their total investment in the external assets). As $D$ increases, banks make some perturbation to the seed and each one approaches a particular asset such that when $D=1$, each bank has invested only in one asset (since all assets are assumed to be the same, it does not matter in which asset a bank specializes. For the perturbation, a vector made of $n$ elements, all of which except one are negative and whose sum is 0, is added to the seed as follows:
  \[
 X =
 \begin{pmatrix}
  s_1+\epsilon_1 & s_2-\epsilon_2 & \cdots & s_n-\epsilon_n \\
  s_1-\epsilon_n & s_2+\epsilon_1 & \cdots & s_n-\epsilon_{n-1} \\
  \vdots  & \vdots  & \ddots & \vdots  \\
  s_1-\epsilon_2 & s_2-\epsilon_3 & \cdots & s_n+\epsilon_1 \\
 \end{pmatrix}
\]
Here $\bigl(\begin{smallmatrix}
s_1 & s_2 & \cdots & s_n 
\end{smallmatrix} \bigr)$ is the initial seed and $\bigl(\begin{smallmatrix}
\epsilon_1 & -\epsilon_2 & \cdots & -\epsilon_n 
\end{smallmatrix} \bigr)$ is the perturbation vector that for each bank is added to the initial seed as above.\par

For such an $X$ matrix, given that $\epsilon_2 \geq \epsilon_3 \geq \cdots \geq \epsilon_n$, $D$ would be:
\begin{equation}
       D=\frac{n[(n-1)(\epsilon_1+\epsilon_2)+\sum_{i=1}^{n-2}{[(n-1-2i)\epsilon_{i+2}]}]}{n(n-1)}  
       \label{eq:diverse}     
    \end{equation}\\
    So, by replacing $\epsilon_1$ with $\epsilon_2+\cdots+\epsilon_n$ in equation \ref{eq:diverse}, there would be:
\begin{equation}
       (2n-2)\epsilon_2+\sum_{i=1}^{n-2}{[(2n-2-2i)\epsilon_{i+2}]}=Dn(n-1)        
       \label{eq:divi}
    \end{equation}
    Therefore, $n-1$ random numbers are generated and after descendingly ordering are passed to $\epsilon_2, \cdots, \epsilon_n$ respectively and these $\epsilon_i$, $i\geq 1$ are normalized so as to satisfy equation \ref{eq:divi}. Since short-selling is not allowed in this model, all the elements of $X$ matrix should be non-negative and if it has not been the case, the above process is repeated until all the elements become non-negative.\par
    For large values of $D$, it is computationally better that the seed be closer to the seed weighting equally all the assets, so it may be better off that the seed feasible area become restricted to some extent corresponding to how large $D$ is.

\end{document}